\documentclass[12pt]{article}
\usepackage{amssymb}
\usepackage{amsfonts}
\usepackage{youngtab}
\usepackage{amsmath,amssymb,latexsym,cite}
\usepackage{amsthm}
\usepackage[]{hyperref}
\input xy
\xyoption{all}

\begin{document}

\vspace*{0.5in}

\begin{center}

{\large\bf Quantum Sheaf Cohomology and Duality of Flag Manifolds}

\vspace{0.2in}

Jirui Guo

\vspace*{0.2in}

\begin{tabular}{c}
Department of Physics and Center for Field Theory and Particle Physics \\
Fudan University, 220 Handan Road, 200433 Shanghai, China
\end{tabular}

{\tt jrguo@fudan.edu.cn}

\end{center}

\hfill \\
\hfill

We study the quantum sheaf cohomology of flag manifolds with
deformations of the tangent bundle and use the ring structure to
derive how the deformation transforms under the biholomorphic
duality of flag manifolds. Realized as the OPE ring of A/2-twisted
two-dimensional theories with (0,2) supersymmetry, quantum sheaf
cohomology generalizes the notion of quantum cohomology. Complete descriptions of quantum sheaf cohomology have been obtained for abelian
gauged linear sigma models (GLSMs) and for nonabelian GLSMs describing Grassmannians. In this paper we continue to explore the quantum sheaf cohomology of nonabelian
theories. We first propose a method to compute the generating
relations for (0,2) GLSMs with (2,2) locus. We apply this method to derive
the quantum sheaf cohomology of products of Grassmannians and flag
manifolds. The dual deformation associated with the biholomorphic duality
gives rise to an explicit IR duality of two A/2-twisted (0,2) gauge
theories.

\newpage

\tableofcontents

\newpage

\section{Introduction}

Quantum cohomology has been an important concept in algebraic
geometry and string theory. It captures nonperturbative corrections
to charged matter couplings in heterotic string compactifications when the gauge connection is determined by the spin connection. In this case, the worldsheet theory has (2,2) supersymmetry. The
coupling constants of the four-dimensional theory can be computed by
the A and B model topological field theories. The OPE ring of the A
model gives rise to quantum cohomology, which has been well-studied in the mathematics literature.

Quantum sheaf cohomology is a generalization of quantum cohomology.
It emerges when the worldsheet theory has only (0,2) supersymmetry. In this case, the
charged matter couplings can be computed by the A/2 and B/2
pseudo-topological field theories \cite{KS}. (See also {\it e.g.} \cite{ S2,S3,MSS,ABS,GK,CGSW,GuS1,GuS2,KMMP,DLM,SM14,ANG1,LU,G1,BBCL,GJS,JSW}.) The moduli space of the (0,2) theory may contain a locus on which supersymmetry is enhanced to (2,2). It was shown in \cite{ADE} that the OPE rings of the A/2 and B/2 models,
which are finite-dimensional truncations of the infinite-dimensional
chiral ring, are still topological at least in a neighborhood of
the (2,2) locus.

Quantum sheaf cohomology is the OPE ring of
the A/2 model, which reduces to the ordinary quantum cohomology on
the (2,2) locus. If the left moving fermions couple to a vector bundle $\mathcal{E}$ over the target space $X$, then the underlying bigraded vector space of this ring is
\[
\bigoplus_{p,q} H^p(X,\wedge^q \mathcal{E}^\vee),
\]
whose product structure encodes the (0,2) generalization of Gromov-Witten invariants. On the moduli space of (0,2) deformations, there are codimension-one degenerate loci, along which the theory does not define a vector bundle or the quantum sheaf cohomology cannot be described as a deformation of the ordinary quantum cohomology. But if the moduli space has a (2,2) locus, then these loci do not intersect the (2,2) locus. We take the (0,2) deformations discussed in this paper to be generic so that they do not lie on the degenerate loci.

Quantum sheaf cohomology on toric varieties has been studied in detail. A general description of the ring structure has been found from both
physical perspectives \cite{MM, MM2} and mathematical perspectives
\cite{DGKS1, DGKS2}. It is shown that quantum sheaf cohomology can be represented by generators and relations.

A first step toward a description of the quantum sheaf cohomology
associated with nonabelian GLSMs is the study of deformed
tangent bundles of Grassmannians. A purely mathematical derivation of the classical sheaf
cohomology of Grassmannians can be found in \cite{GLS2}. Quantum corrections are taken into account in \cite{GLS1} by using the relations encoded in the one-loop effective
potential.

In this paper, we first propose a method to compute the ring relations of quantum sheaf cohomology for any (0,2) GLSM which has a (2,2) locus on its moduli space. Though the one-loop effective potential on the Coulomb branch encodes the quantum relations, it does not contain enough information for the purpose of writing down all the relations in a gauge invariant way. So our idea is to make use of the localization formula of \cite{CGJS} to deduce the classical relations and then use the one-loop relations to incorporate quantum corrections. (See {\it e.g.} \cite{BL,CMP,CCP} for more discussions on localization.) We describe this method in detail in section \ref{section:method}.

We first apply this method to Grassmannians to reproduce the result of \cite{GLS1, GLS2} and generalize it to the direct product of an arbitrary number of Grassmannians. We then use this method to study the quantum sheaf cohomology of deformed
tangent bundles of flag manifolds. The flag manifold
$F(k_1,k_2,\cdots,k_n,N)$ can be described by the geometric phase of
an $N=(2,2)$ GLSM with
\[
U(k_1)\times U(k_2) \times \cdots \times U(k_n)
\]
gauge group, one chiral multiplet in the bifundamental representation of $U(k_i)$ and $U(k_{i+1})$ for each $i$ and $N$ chiral multipliets in the fundamental representation of $U(k_n) $\cite{DS}. We deform the theory
by turning on nontrivial $E$-terms and thus break the (2,2)
supersymmetry to (0,2) supersymmetry. The deformation of the $E$-terms is
parameterized by a set of constants $u^s_t$ and a set of $N \times
N$ matrices $A_t$ for $t=1,\cdots,n,~s=1,\cdots,n-1$. It turns out that the quantum sheaf cohomology of flag manifolds can be written as
\[
\mathbb{C}[x^{(1)}_1, x^{(1)}_2 \cdots, x^{(2)}_1,\cdots,
x^{(n+1)}_1, \cdots]/(I+R),
\]
where $I$ and $R$ are ideals of relations to be computed in later
sections. In this representation, only the ideal $R$ depends on the (0,2) deformation given by the $E$-terms.

There is a biholomorphic duality between $F(k_1,k_2,\cdots,N)$ and
$F(N-k_n,N-k_{n-1},\cdots,N-k_1,N)$. When there is a deformation on
$F(k_1,k_2,\cdots,N)$ given by the data $u^s_t$ and $A_t$, there
should be a dual deformation on $F(N-k_n,N-k_{n-1},\cdots,N-k_1,N)$
given by ${u'}^s_t$ and $A'_t$ such that the deformed tangent
bundles coincide. A natural question to ask is how the two sets of
data are related. If we can express ${u'}^s_t$ and $A'_t$ as
functions of $u^s_t$ and $A_t$, we get a pair of (0,2) GLSMs in IR
duality, an analogue of Seiberg duality. One side of the duality is a $U(k_1)\times U(k_2) \times \cdots \times U(k_n)$ quiver gauge theory, the other is a $U(N-k_n)\times U(N-k_{n-1}) \times \cdots \times U(N-k_1)$ quiver gauge theory. The same problem
can be studied for product of Grassmannians. We will solve this
problem in section \ref{section:dual} by using ring relations of the
quantum sheaf cohomology.

This paper is organized as follows. In section \ref{section:general} we first discuss general issues regarding $(0,2)$ deformation of GLSMs and then address our algorithm computing the ring structure of quantum sheaf cohomology corresponding to $(0,2)$ theories with (2,2) locus. We apply this method to deformed tangent bundles of Grassmannians and products of multiple Grassmannians. We study general flag manifolds in section \ref{section:FLAG}. We first review the structure of ordinary quantum cohomology and rewrite it in a form that is suitable for generalizing to the (0,2) case. Then we use our method to compute the quantum sheaf cohomology of flag manifolds. In section \ref{section:dual}, we derive the explicit relationship between the (0,2) deformations associated with the biholomorphic duality of flag manifolds. This correspondence amounts to an analogue of Seiberg duality of (A/2-twisted) (0,2) quiver gauge theories.

\section{Quantum sheaf cohomology of (0,2) GLSMs}\label{section:general}

\subsection{(0,2) deformation}

As mentioned in the introduction, we only consider $\mathcal{N}=(0,2)$ GLSMs
with $\mathcal{N}=(2,2)$ loci. Since our interests lie in quantum sheaf
cohomology, which does not depend on nonlinear
deformations, we can assume that the theory under consideration is
defined by some linear deformation of the $E$-terms of an $\mathcal{N}=(2,2)$
theory. More precisely, let's consider a GLSM with gauge group $G$
and corresponding Lie algebra $\mathfrak{g}$. The theory
consists of a $\mathfrak{g}$-valued vector multiplet, a chiral
multiplet $\Sigma$ in the adjoint representation of $\mathfrak{g}$,
chiral multiplets $\Phi_i$ and Fermi multiplets $\Lambda_i$. For
each $i$, $\Phi_i$ and $\Lambda_i$ are in the same representation
$\mathcal{R}_i$ of $\mathfrak{g}$. The $E$-functions are given by
\[
\overline{D}_+ \Lambda^i = E^i (\Sigma, \Phi^j),
\]
where $E^i$ is a holomorphic function linear in $\Sigma$ and in each
$\Phi^j$. On the (2,2) locus, the $E$ functions are given by
\[
E^i = \Sigma \Phi^i,
\]
where $\Sigma$ acts on $\Phi^i$ according to the representation
$\mathcal{R}_i$.

For example, the Grassmannian $G(k,N)$ is described by a $U(k)$
gauge theory with $N$ chiral multiplets and $N$ fermi multiplets in
the fundamental representation of $U(k)$. Then the $E$-functions with linear deformation are given by
\begin{equation}\label{E_G}
\overline{D}_+ \Lambda^i_\alpha = \Sigma^\beta_\alpha
\Phi^i_\beta + A^i_j (\mathrm{Tr} \Sigma) \Phi^j_\alpha
\end{equation}
up to a field redefinition, where $\alpha,\beta$ are gauge indices and $A$ is an $N \times N$ matrix. If we take $\mathcal{S}$ to be the universal bundle and $\mathcal{V}$ to be the trivial bundle of rank $N$ over $G(k,N)$, then the deformation given above defines a deformed tangent bundle through the short exact sequence
\begin{equation}\label{ses_G}
0 \to \mathcal{S} \otimes \mathcal{S}^\vee \stackrel{g}{\to}
\mathcal{V} \otimes \mathcal{S}^\vee \to \mathcal{E} \to 0.
\end{equation}
The above map $g$ can be represented as
\[
\sigma^{\beta}_{\alpha} \mapsto \sigma^{\beta}_{\alpha}
x^i_{\beta} + \sigma^{\beta}_{\beta} A^i_j x^j_{\alpha},
\]
where $x$ is the homogeneous coordinate of the Grassmannian.
On the (2,2) locus $A$ vanishes and $\mathcal{E}$ becomes the tangent
bundle. For special values of $A$, the deformation does not define a vector bundle or the quantum sheaf cohomology is not a deformation of the ordinary quantum cohomology. These values form codimension-one subvarieties, called the degenerate loci, of the whole (0,2) moduli space, see \cite{GLS1, GLS2} for more details.

For the theories studied in this paper, we only consider the generic case, i.e. we assume the (0,2) deformation is not on a degenerate locus of the moduli space.

\subsection{Ring structure}\label{section:method}

The ring structure of quantum sheaf cohomology can be derived from the OPE on the Coulomb branch. We denote the scalar component of $\Sigma$ by $\sigma$. On the Coulomb
branch, the operators in the quantum sheaf
cohomology correspond to gauge invariant polynomials in the diagonal elements of
$\sigma$. For example, for the Grassmannian $G(k,N)$, because the
Weyl group is the permutation group of $k$ elements, the operators
in the quantum sheaf cohomology are symmetric polynomials in the
diagonal elements of $\sigma$.

For Grassmannians, the classical relations have been derived using purely
algebro-geometric techniques in \cite{GLS2}, and quantum corrections
are computed in \cite{GLS1} by use of the one-loop effective
potential on the Coulomb branch.

In this paper we are going to study more complicated nonabelian
gauge theories and we will adopt another approach. This approach makes use of the localization formula of \cite{CGJS} and equations of motion derived from the one-loop effective potential. We now discuss the idea of this approach.

On the Coulomb branch, the $E$-functions read
\[
E_i = \sigma_a E^a_i (\phi),
\]
where $E^a_i$ are holomorphic functions. Then the mass matrix of the
matter multiplets is
\[
\left. M_{ij} = \frac{\partial E_i}{\partial \phi_j}
\right|_{\phi=0}.
\]
Gauge invariance and the $U(1)_R$ symmetry imply that the matrix $M$ is
block diagonal, with each block mixing fields with the same
$R$-charge and the same weight of the gauge symmetry. Let us denote
by $M_{(\gamma,\rho_\gamma)}$ the block associated with the weight
$\rho_\gamma$ of the representation $\gamma$ of the gauge group. We
denote by $r_\gamma$ the $R$-charge of the corresponding chiral
multiplet\footnote{Then the corresponding fermi multiplet has $R$-charge
$r_\gamma-1$.}. For the theories we consider in this paper, $r_\gamma
= 0$. The common zero set of all the $M_{(\gamma,\rho_\gamma)}$'s is the origin.

The localization formula of \cite{CGJS} expresses the $A/2$-twisted
correlation functions on the sphere as a sum over flux sectors with
each summand given by a Jeffrey-Kirwan-Grothendieck residue.
Explicitly, for any operator $\mathcal{O}(\sigma)$ written as a
gauge invariant polynomial, the correlation function is
\begin{equation}\label{localization}
\langle \mathcal{O}(\sigma) \rangle = \frac{(-1)^{N_*}}{|W|} \sum_{k
\in \Gamma_{G^\vee}} q^k \mathrm{JKG-Res}[\eta]
\mathcal{Z}^{1-\mathrm{loop}}_k (\sigma) \mathcal{O}(\sigma)
d\sigma_1 \wedge \cdots \wedge d\sigma_{\mathrm{rk}(G)},
\end{equation}
where
\begin{equation}\label{z_oneloop}
\mathcal{Z}^{1-\mathrm{loop}}_k =
(-1)^{\sum_{\alpha>0}(\alpha(k)+1)} \prod_{\alpha>0}
\alpha(\sigma)^2 \prod_{\gamma} \prod_{\rho_\gamma \in R_\gamma}
\left( \det (M_{(\gamma,\rho_\gamma)})
\right)^{r_\gamma-1-\rho_\gamma(k)}.
\end{equation}
Here $W$ is the Weyl group of the gauge group. $N_*$ is the
number of chiral multiplets with $R$-charge 2. $\Gamma_{G^\vee}
\cong \mathbb{Z}^{\mathrm{rk}(G)}$ is the dual lattice of the weight
lattice and
\[
q^k \equiv  \exp(2 \pi i \tau^a k_a)
\]
with $\tau^a$ the complexified FI parameter. In \eqref{z_oneloop},
the first product is taken over all the positive roots of the gauge
group. We define
\[
\Delta \equiv \prod_{\alpha>0} \alpha(\sigma).
\]

In \eqref{localization}, $\mathrm{JKG-Res}[\eta]$ refers to the
Jeffrey-Kirwan-Grothendieck residue, which depends on an element
$\eta$ in the dual space of the Cartan subalgebra of the gauge
group. The correlation function does not depend on the choice of
$\eta$. As in \cite{CGJS}, we take $\eta$ to be the effective FI
parameter at infinity on the Coulomb branch:
\begin{equation}\label{eta}
\eta = \xi + \frac{1}{2 \pi} b_0 \lim_{\Lambda \to \infty} \log \Lambda,
\end{equation}
where $\xi$ is the FI parameter appearing in the FI term of the UV
GLSM and
\begin{equation}\label{b0}
b_0 = \sum_i \sum_{\rho_i \in R_i} \rho_i,
\end{equation}
where $i$ runs over all the chiral multiplets and $\rho_i$ runs over
all the weights of the represenation $\mathcal{R}_i$. For example, $b_0^a = N$
for all $a = 1,\cdots, k$ in the Grassmannian case described in the
last subsection.

An operator $\mathcal{O}_R$ is zero in the quantum sheaf cohomology if
and only if the $A/2$ correlation function
\begin{equation}\label{corr}
\langle \mathcal{O}_R \mathcal{O} \rangle = 0
\end{equation}
for any operator $\mathcal{O}$. We will first derive the classical relations, so we take the $q \to 0$ limit in \eqref{localization} and thus keep only the $k=0$ term. As we will see, for the theories we consider in this paper, $\eta$ lies in the cone spanned by all the weights
$\rho_\gamma$ appearing in \eqref{z_oneloop}, so the JKG-residue
reduces to the ordinary Grothendieck residue at $\sigma=0$. Because $r_\gamma=0$,
\eqref{corr} implies
\[
\mathrm{Res}_{(0)} \frac{\Delta^2 \mathcal{O}_R \mathcal{O}}{\prod_{\gamma} \prod_{\rho_\gamma \in R_\gamma}
\left( \det (M_{(\gamma,\rho_\gamma)})
\right)} d\sigma_1 \wedge \cdots \wedge d\sigma_{\mathrm{rk}(G)} = 0
\]
for any operator $\mathcal{O}$.
Thus $\Delta^2 \cdot \mathcal{O}_R$ must lie in the ideal
generated by $\det (M_{(\gamma,\rho_\gamma)})$ with $\gamma$ and
$\rho_\gamma$ running over all the representations and weights of
the matter multiplets due to the property of Grothendieck residues \footnote{See (0.7) of \cite{CDS}.}.

In general, the quantum sheaf cohomology of the theories we consider
is of the form
\[
\mathcal{A}/(I+R),
\]
where $\mathcal{A}$ is a free $\mathbb{C}$-algebra, $I$ is an ideal
independent of the (0,2) deformation while the ideal $R$ depends on
the deformation.

Assuming we know the ideal $R$ on the (2,2) locus, the relations off
the (2,2) locus can be found via the following procedure. First, we
turn off the (0,2) deformation. Let us denote the mass matrices on the (2,2) locus by
$M^{(2,2)}_{(\gamma,\rho_\gamma)}$. From the discussion above, the (2,2) relations in
$R$ multiplied by $\Delta^2$ can be expressed as polynomials in $\det
(M^{(2,2)}_{(\gamma,\rho_\gamma)})$. Let $R^{(2,2)}_r$ be a set of relations
generating $R$ on the (2,2) locus, then we can
write
\[
\Delta^2 R^{(2,2)}_r = P_r(\det (M^{(2,2)}_{(\gamma,\rho_\gamma)}))
\]
for some polynomials $P_r$.

Next we turn on (0,2) deformation
and use the expressions for $\det (M_{(\gamma,\rho_\gamma)})$ with generic deformations to substitute $\det (M^{(2,2)}_{(\gamma,\rho_\gamma)})$:
\[
\Delta^2 R_r = P_r(\det (M_{(\gamma,\rho_\gamma)})).
\]
Once we rewrite the relations $R_r$ above in a gauge invariant
way, we get a set of generators of $R$ in the classical limit for generic (0,2) deformation. Note that
this is guaranteed because the deformation is not on the degenerate locus.

At a generic point on the Coulomb branch, the effective $J$-functions are \cite{CGJS, MM, MM2}
\[
J^a = \tau^a - \frac{1}{2 \pi i} \sum_\gamma \sum_{\rho_\gamma \in R_\gamma} \rho^a_\gamma \log \left( \det (M_{(\gamma,\rho_\gamma)}) \right) - \frac{1}{2}\sum_{\alpha>0} \alpha^a,
\]
where $a$ is the index of the Cartan subalgebra.
Quantum corrections to the relations can be incorporated by using the equations of
motion derived from the effective $J$-functions above, which read
\begin{equation}\label{EOM_general}
\prod_{\gamma} \prod_{\rho_\gamma} \left(\det
(M_{(\gamma,\rho_\gamma)})\right)^{\rho_\gamma^a} = (-1)^{\sum_{\alpha
> 0} \alpha^a} q_a
\end{equation}
for all $a$. From these equations one can compute the quantum relations.

\subsection{Grassmannians}

In order to illustrate the idea, let's follow the procedure
described in the last subsection to reproduce the quantum sheaf
cohomology of the Grassmannian $G(k,N)$ found in \cite{GLS1, GLS2}.

Let's denote the diagonal elements of $\sigma$ by $\sigma_a,
a=1,\cdots,k$. We have seen the operators are given by symmetric
polynomials in $\sigma_a$. There are various choices for a
basis of the symmetric polynomials, we could in principle take any
choice as a basis of the quantum sheaf cohomology. It turns
out that the Schur polynomials provide a natural choice due to their
geometric interpretation as Schubert classes.

The Schur polynomials in $k$ variables are labeled by dominant
integral weights of $\mathfrak{u}(k)$, which correspond to partitions of integer into at
most $k$ parts. We will denote by $S_{\mu}$ the Schur polynomial
associated with the weight $\mu$. Such a weight can be written as
$\mu=(\mu_1,\mu_2,\cdots,\mu_s)$ where the $\mu_i$'s form a sequence
of nonincreasing integers. We will use $(1^s)$ to denote the
partition of $s$ into $s$ parts, i.e.
\[
(1^s) = (\underbrace{1,1,\cdots,1}_{s}).
\]
The ordinary cohomology ring of $G(k,N)$ can be written as
\begin{equation}\label{ring0_G}
\mathbb{C}[x_1,x_2,\cdots,y_1,y_2,\cdots]/(I+R).
\end{equation}
The ideal $R$ is generated by
\[
\{x_i, y_j ~|~ i>k, j>N-k\},
\]
and $I$ is generated by
\[
\left\{\sum_{i+j=m} x_i y_j~ | ~ m>0 \right\}.
\]
Of course we could use the relations of $I$ to express $x$ as polynomials of $y$, which read
\[
x_m = (-1)^m \det(y_{1+j-i})_{1\leqslant i,j\leqslant m},
\]
and thus reproduce the usual representation \cite{buch}.
The geometric meaning of these relations becomes transparent once we
identify $x_i$ with the $i$-th Chern class of the universal bundle
$\mathcal{S}$ and $y_j$ with the $j$-th Chern class of the universal
quotient bundle $\mathcal{Q}$, where $\mathcal{S}$ and $\mathcal{Q}$ are
related through the Euler sequence
\[
0 \to \mathcal{S} \to \mathcal{V} \to \mathcal{Q} \to 0.
\]
As in \cite{GLS1}, by interpreting the $\sigma_a$'s as Chern roots
of $\mathcal{S}^\vee$, we get
\begin{equation}\label{xy_G}
\begin{split}
&x_i=(-1)^i S_{(1^i)}(\sigma), \\
&y_j=S_{(j)}(\sigma).
\end{split}
\end{equation}
The vanishing of $x_i$ for $i>k$ is merely a consequence of the fact
that there are $k$ variables. In order to apply our method discussed in section \ref{section:method}, we need to express the relations $y_j$ multiplied by $\Delta^2$ as polynomials of $\det(M^{(2,2)}_a)$. For that purpose, we need the following
definition of Schur polynomials
\begin{equation}\label{schur}
S_{\lambda}(\sigma) =
\frac{\det(\sigma^{\lambda_i+k-i}_j)_{1\leqslant i,j \leqslant
k}}{\Delta},
\end{equation}
where $\Delta = \prod_{a<b}(\sigma_a-\sigma_b)$ ($\Delta = 1$ when
$k=1$). In the current situation, $\det(M_a^{(2,2)})=\sigma_a^N$ on the
(2,2) locus. Using \eqref{schur} and expanding the determinant along
the first row, we get
\begin{equation}\label{G(k,N)reln(2,2)}
\Delta^2 y_{N-k+r} = \Delta \sum_{a=1}^{k} (-1)^{a-1} \det(M_a^{(2,2)})
\sigma^{r-1}_a \theta_a,
\end{equation}
where
\[
\theta_a= \prod_{i<j \atop i,j\neq a}(\sigma_i-\sigma_j)
\]
for $k>2$ and $\theta_a=1$ for $k=1,2$. Note that
\begin{equation}\label{xita}
\sum_{a=1}^k (-1)^{a-1}\sigma_a^m \theta_a = \left\{
\begin{array}{ll}
0, & 0 \leq  m \leq k-2, \\
S_{(m-k+1)}(\sigma) \Delta, & m \geq k-1.
\end{array} \right.
\end{equation}

Off the (2,2) locus, the relations $x_i$ for $i>k$ are unchanged due
to \eqref{xy_G}. Assume that the relations $y_{N-k+r}$ for $r>0$
are deformed to $R_{N-k+r}$ off the (2,2) locus. To compute $R_{N-k+r}$ we use the (0,2) expression of
$\det(M_a)$ to substitute $\det(M_a^{(2,2)})$ in \eqref{G(k,N)reln(2,2)}. If we use $I_i(\Omega)$ to denote the $i$-th characteristic polynomial of a matrix $\Omega$, then
\[
\det(M_a)=\det(\sigma_aI + y_1 A)=\sum_{i=0}^{N} \sigma^{N-i}_a
I_i(y_1 A).
\]
From \eqref{G(k,N)reln(2,2)} and \eqref{xita}, we get
\begin{equation}\label{G(k,N)R}
\begin{split}
R_{N-k+r} = &\Delta^{-1} \sum_{a=1}^k (-1)^{a-1} \det(M_a)
\sigma^{r-1} \theta_a \\
=& \sum_{i=0}^{\min\{N,N-k+r\}}I_i(y_1 A)y_{N-k+r-i}.
\end{split}
\end{equation}
Then the ideal $R$ in \eqref{ring0_G} is generated by
\[
\{ R_r~ |~ r>N-k\}
\]
in the classical limit.

To get the quantum relations, we use the identity
\[
\det(M_a)+q=0,
\]
derived from the one-loop effective $J$-function
\[
J_a = - \ln\left[ -q^{-1} \det( M_a )\right].
\]
Here we treat $q$ as a degree $N$ element in the quantum sheaf
cohomology. From \eqref{G(k,N)R} and \eqref{xita}, we get
\[
\Delta^2 R_{N-k+r}= -q \Delta \sum_{a=1}^k (-1)^{a-1} \sigma^{r-1}
\theta_a = \left\{
\begin{array}{ll}
0, & 1\leq r\leq k-1, \\
-q \Delta^2 S_{(r-k)}(\sigma), & r \geq k.
\end{array} \right.
\]
In view of \eqref{xy_G}, the identity above means
\[
R_{N-k+r}+qy_{r-k}=0
\]
for $r \geq k$ in the quantum sheaf cohomology ring.
So if we define
\[
\tilde{R}_{r} = \left\{
\begin{array}{ll}
R_{r}, & r<N, \\
R_{r} + q y_{r-N}, & r \geqslant N,
\end{array} \right.
\]
then the ideal $R$ is generated by
\[
\{\tilde{R}_r ~|~ r>N-k\}.
\]
This is in agreement with the result of \cite{GLS1, GLS2}.

\subsection{Product of Grassmannians}\label{section:GG}

In this subsection we use our method to derive a new result,
namely the quantum sheaf cohomology on direct product of multiple
Grassmannians. We start with the product of two Grassmannians.

The direct product of two Grassmannians $G(k_1,N_1) \times
G(k_2,N_2)$ can be realized by a $U(k_1)\times U(k_2)$ gauge theory.
On the (2,2) locus, the $U(k_1)$ and $U(k_2)$ sectors decouple,
there are $N_1$ chiral multiplets in the fundamental
representation of $U(k_1)$ and $N_2$ chiral multiplets in the
fundamental representation of $U(k_2)$. In (0,2) language, we denote
by $\Phi_s$ and $\Lambda_s$ the chiral and fermi multiplets in the
fundamental representation of $U(k_s)$ and by $\Sigma_s$ the chiral
multiplet in the adjoint representation of $U(k_s)$ for $s=1,2$. Off the (2,2) locus, the two sectors are coupled by
$E$-terms:
\begin{equation}\label{E_GG}
\begin{split}
&\overline{D}_+ {\Lambda_1}^{i_1}_a = {\Phi_1}^{i_1}_b
{\Sigma_1}^b_a + ({\Sigma_1}^b_b
A^{i_1}_{j_1}+{\Sigma_2}^\beta_\beta
B^{i_1}_{j_1}) {\Phi_1}^{j_1}_a,\\
&\overline{D}_+ {\Lambda_2}^{i_2}_\alpha = {\Phi_2}^{i_2}_\beta
{\Sigma_2}^\beta_\alpha + ({\Sigma_1}^b_b
C^{i_2}_{j_2}+{\Sigma_2}^\beta_\beta D^{i_2}_{j_2})
{\Phi_2}^{j_2}_\alpha
\end{split}
\end{equation}
for $i_1,j_1=1,\cdots,N_1, i_2,j_2=1,\cdots,N_2, a,b=1,\cdots,k_1,
\alpha,\beta=1,\cdots,k_2$.

Define
\[
\begin{split}
&M_{1a} = \sigma_{1a}I + (\mathrm{Tr}\sigma_1) A +
(\mathrm{Tr}\sigma_2) B, \\
&M_{2\alpha} = \sigma_{2\alpha}I + (\mathrm{Tr}\sigma_1) C +
(\mathrm{Tr}\sigma_2) D,
\end{split}
\]
then
\begin{equation}\label{detM_GG}
\begin{split}
&\det(M_{1a})=\sum_{i=0}^{N_1} \sigma_{1a}^{N_1-i}
I_i((\mathrm{Tr}\sigma_1) A + (\mathrm{Tr}\sigma_2) B)\\
&\det(M_{2\alpha})=\sum_{j=0}^{N_2} \sigma_{2\alpha}^{N_2-j}
I_j((\mathrm{Tr}\sigma_1) C + (\mathrm{Tr}\sigma_2) D).
\end{split}
\end{equation}
On the (2,2) locus, $A=B=C=D=0$ and $\det(M^{(2,2)}_{1a})={\sigma_1}^{N_1}_a,
\det(M^{(2,2)}_{2\alpha})={\sigma_2}^{N_2}_\alpha$. Since $b_0^a=N_1$ for
$a=1,\cdots,k_1$ and $b_0^a=N_2$ for $a=k_1+1,\cdots,k_2$, we see
$\eta$ defined in \eqref{eta} is indeed in the cone spanned by all
the weights of the representation of the chiral multiplets.

The classical cohomology of $G(k_1,N_1) \times G(k_2,N_2)$ is
\begin{equation}\label{ring0_GG}
\mathcal{A}_{G(k_1,N_1) \times G(k_2,N_2)} =
\mathbb{C}[x_{11},x_{12},\cdots,y_{11},y_{12},\cdots,x_{21},x_{22},\cdots,y_{21},y_{22},\cdots]/(I+R),
\end{equation}
where $I$ is generated by
\[
\left\{ \sum_{i_1+j_1=m_1} x_{1i_1} y_{1j_1},~\sum_{i_2+j_2=m_2} x_{1i_2}
y_{1j_2} ~|~ m_1>0,~m_2>0 \right\}
\]
and $R$ is generated by
\[
\{ x_{1 i_1},~ y_{1 j_1},~ x_{2 i_2},~ y_{2 j_2} ~|~ i_1>k_1,~
j_1>N_1-k_1,~ i_2>k_2,~ j_2>N_2-k_2 \}.
\]
Define
\[
\Delta_1=\prod_{a<b} (\sigma_{1a}-\sigma_{1b}),\quad
\Delta_2=\prod_{\alpha<\beta} (\sigma_{2\alpha}-\sigma_{2\beta}).
\]
$x_{si}$ and $y_{sj}$ are identified with the $i$-th Chern class of
the universal bundle and the $j$-th Chern class of the universal
quotient bundle of $G(k_s,N_s)$ respectively, i.e.
\begin{equation}\label{xy_GG}
\begin{split}
&x_{si}=(-1)^i S_{(1^i)}(\sigma_s), \\
&y_{sj}=S_{(j)}(\sigma_s).
\end{split}
\end{equation}
As in the case of a single Grassmannian, we have
\begin{equation}\label{(2,2)relation_GG}
\Delta_s y_{N_s-k_s+r} = \Delta_s S_{(N_s-k_s+r)}(\sigma_s)
=\sum_{a=1}^{k_s} (-1)^{a-1} \det(M^{(2,2)}_{sa}) {\sigma_s}^{r-1}_a
\theta_{sa}
\end{equation}
for $s=1,2$ on the (2,2) locus, where $\theta_{sc} = \prod_{a<b
\atop a,b\neq c} (\sigma_{sa}-\sigma_{sb})$. Now we turn on (0,2)
deformation by assigning nonzero values to the matrices
$A, B, C$ and $D$. Define
\[
\begin{split}
R_{1r}=\sum_{i=0}^{\min\{r,N_1\}} I_i(y_{11} A + y_{21} B) y_{1,r-i},\\
R_{2r}=\sum_{i=0}^{\min\{r,N_2\}} I_i(y_{11} C + y_{21} D) y_{2,r-i}
\end{split}
\]
for $r\geq 0$. From \eqref{detM_GG} and \eqref{xy_GG}, the
right hand side of \eqref{(2,2)relation_GG} becomes
\begin{equation}\label{relation_GG}
\begin{split}
\sum_{a=1}^{k_1} (-1)^{a-1} \det(M_{1a}) {\sigma_1}^{r-1}_a
\theta_{1a} = \Delta_1 \sum_{i=0}^{N_1-k_1+r} I_i(y_{11} A + y_{21}
B) y_{1,N_1-k_1+r-i},\\
\sum_{\alpha=1}^{k_2} (-1)^{\alpha-1} \det(M_{2\alpha})
{\sigma_2}^{r-1}_\alpha \theta_{2\alpha} = \Delta_2
\sum_{i=0}^{N_2-k_2+r} I_i(y_{11} C + y_{21} D) y_{2,N_1-k_1+r-i}.
\end{split}
\end{equation}
Then \eqref{(2,2)relation_GG} implies that the classical sheaf
cohomology is given by \eqref{ring0_GG} with $R$ generated by
\[
\{ x_{1 i_1},~ R_{1 j_1},~ x_{2 i_2},~ R_{2 j_2} ~|~ i_1>k_1,~
j_1>N_1-k_1,~ i_2>k_2,~ j_2>N_2-k_2 \}.
\]
Thus we have obtained the representation of classical sheaf cohomology on $G(k_1,N_1) \times G(k_2,N_2)$.

Now we turn to the quantum case. Vanishing of the effective $J$-functions implies
\begin{equation}\label{emo_GG}
\det(M_{1a}) = -q_1,~\det(M_{2\alpha}) = -q_2.
\end{equation}
Again, to deduce the quantum relations, we simply replace
$\det(M_{1a})$ and $\det(M_{2\alpha})$ on the left hand side of
\eqref{relation_GG} using \eqref{emo_GG} and employ \eqref{xita} to get
\[
\tilde{R}_{1r} = \tilde{R}_{2t}=0,~~ r>N_1-k_1, t>N_2-k_2,
\]
where
\[
\tilde{R}_{1r} = \left\{
\begin{array}{ll}
R_{1r}, & r<N_1, \\
R_{1r} + q_1 y_{r-N_1}, & r \geqslant N_1,
\end{array} \right.
\]
\[
\tilde{R}_{2t} = \left\{
\begin{array}{ll}
R_{2t}, & t<N_2, \\
R_{2t} + q_2 y_{t-N_2}, & t \geqslant N_2.
\end{array} \right.
\]
Note that $\mathrm{deg}(q_1)=N_1, \mathrm{deg}(q_2)=N_2$. We
conclude that the quantum sheaf cohomology of $G(k_1,N_1) \times
G(k_2,N_2)$ is
\begin{equation}\label{ring_GG}
\mathcal{A}_{G(k_1,N_1) \times G(k_2,N_2)} =
\mathbb{C}[x_{11},x_{12},\cdots,y_{11},y_{12},\cdots,x_{21},x_{22},\cdots,y_{21},y_{22},\cdots]/(I+R),
\end{equation}
where $I$ is the same as the (2,2) case and $R$ is generated by
\[
\{ x_{1 i_1},~ \tilde{R}_{1 j_1},~ x_{2 i_2},~ \tilde{R}_{2 j_2} ~|~
i_1>k_1,~ j_1>N_1-k_1,~ i_2>k_2,~ j_2>N_2-k_2 \}.
\]

Following the same procedure, the result can be generalized to the direct
product of arbitrarily many Grassmannians $G(k_1,N_1) \times
G(k_2,N_2) \times \cdots \times G(k_n, N_n)$ with the (0,2)
deformation given by
\begin{equation}\label{E_GGGG}
\overline{D}_+ \Lambda_s = \Phi_s \Sigma_s + \sum_{t=1}^n
(\mathrm{Tr}\Sigma_t) A_{st} \Phi_s, ~~s=1,\cdots,n,
\end{equation}
where $A_{st}$ is an $N_s \times N_s$ matrix for every $t$. The generators of the
quantum sheaf cohomology are $x_{si_s}, y_{sj_s}$ for $s=1,\cdots,n,
i_s>0, j_s>0$. The ring can be written as
\begin{equation}\label{ring_Gs}
\mathcal{A}_{G(k_1,N_1) \times G(k_2,N_2) \times \cdots \times
G(k_n, N_n)} = \mathbb{C}[x_{1i_1}, y_{1i_1}
x_{2i_2},y_{2i_2},\cdots, x_{n i_n}, y_{n i_n}]/(I+R)
\end{equation}
for $i_s \geq 1$, where $I$ is generated by
\[
\left\{ \sum_{i_s+j_s=m_s} x_{si_s} y_{sj_s} ~|~ s=1,\cdots,n, m_s >0
\right\}
\]
and $R$ is generated by $\tilde{R}_{s r_s}$ for $s=1,\cdots,n, r_s
> N_s-k_s$, where
\[
\tilde{R}_{s r_s} = \left\{
\begin{array}{ll}
\sum_{i=0}^{\min\{r_s,N_s\}} I_i(\sum_{t=1}^n y_{t1}A_{st}) y_{s,r-i}, & r<N_s, \\
\sum_{i=0}^{\min\{r_s,N_s\}} I_i(\sum_{t=1}^n
y_{t1}A_{st}) y_{s,r-i} + q_s y_{s,r-N_s}, & r
\geqslant N_s.
\end{array} \right.
\]

\section{Flag manifolds}\label{section:FLAG}

For every sequence of integers $(k_1,k_2,\cdots,k_n)$ with
$0 < k_1 < k_2 < \cdots < k_n < N$, the flag
manifold $F(k_1,k_2,\cdots,k_n,N)$ is defined by the set of flags in
$\mathbb{C}^N$
\[
F(k_1,k_2,\cdots,k_n,N) = \{(V_{k_1},\cdots,V_{k_n}) \in G(k_1,N)
\times \cdots \times G(k_n,N) | V_{k_1} \subset \cdots \subset
V_{k_n}\}.
\]
The (2,2) GLSM describing $F(k_1,k_2,\cdots,k_n,N)$ is a quiver gauge
theory with gauge group $U(k_1) \times \cdots \times U(k_n)$
\cite{DS}. For each $s=1,\cdots,n-1$, there is a chiral multiplet
$\Phi_{s,s+1}$ transforming in the fundamental representation of
$U(k_s)$ and in the antifundamental representation of $U(k_{s+1})$.
There are also chiral multiplets $\Phi_{n,n+1}^i$ transforming in
the fundamental representation of $U(k_n)$, $i=1,\cdots,N$. The quiver diagram of this theory is
\[
\xymatrix@C=3pc{
*++[o][F]{k_1} & *++[o][F]{k_2} \ar[l]_{\Phi_{12}} & \cdots \ar[l]_{\Phi_{23}} & *++[o][F]{k_n} \ar[l]_{\Phi_{n-1,n}} & *++[F]{N} \ar[l]_{\Phi_{n,n+1}}
}
\]
There is a flag of universal subbundles
\[
0=\mathcal{S}_0 \hookrightarrow \mathcal{S}_1 \hookrightarrow
\mathcal{S}_2 \hookrightarrow \cdots \hookrightarrow \mathcal{S}_n
\hookrightarrow \mathcal{S}_{n+1}=\mathcal{O}^{\oplus N},
\]
where the fibers of these bundles at any point of the flag manifold
form the flag corresponding to that point, so $\mathcal{S}_i$ has
rank $k_i$. We can regard $\Phi_{s,s+1}$ as the inclusion map
$\mathcal{S}_s \to \mathcal{S}_{s+1}$. The conjugate of
$\Phi_{s,s+1}$ can be viewed as the dual map $\mathcal{S}_{s+1}^\vee \to
\mathcal{S}_{s}^\vee$.

The tangent bundle $TF$ of the flag manifold can be described by the
following short exact sequence
\begin{equation}\label{ses_F}
0 \to \bigoplus_{i=1}^n \mathcal{S}^\vee_i \otimes \mathcal{S}_i
\stackrel{g}{\to} \bigoplus_{i=1}^n \mathcal{S}^\vee_i \otimes
\mathcal{S}_{i+1} \to TF \to 0.
\end{equation}
The diagonal of $g$ consists of the inclusion maps
$\mathcal{S}^\vee_i \otimes \mathcal{S}_i \to \mathcal{S}^\vee_i
\otimes \mathcal{S}_{i+1}$, and the subdiagonal above the diagonal consists of the dual maps $\mathcal{S}^\vee_{i+1} \otimes
\mathcal{S}_{i+1} \to \mathcal{S}^\vee_{i} \otimes
\mathcal{S}_{i+1}$. The deformed tangent bundle is defined by
deforming the map $g$ in \eqref{ses_F}. Before discussing the
deformation, let's first look at the structure of the ordinary
cohomology.

\subsection{Quantum cohomology}

In this subsection we describe the structure of the ordinary quantum
cohomology of flag manifolds. We will give a representation
suitable for generalization to the (0,2) case. (See \cite{BCKS,W,bertram1,AS,CF,K} for more details on quantum cohomology of flag manifolds.)

First, we need to introduce some notations. All the cohomology rings
we talk about in this paper are actually graded algebras. For a set
of homogeneous elements $x_s$ with $\mathrm{deg}(x_s)=s,
s=0,1,2,\cdots$, and $x_0=1$ we write
\[
[x]\equiv \sum_{s=0}^{\infty} x_s.
\]
Sometimes, a sequence $x_s$ terminates at finite degree $s=s_0$, we still use the
definition above by assuming $x_s=0$ for $s>s_0$. The equality
\[
[x]=[y]
\]
means $x_s = y_s$ for all $s\geq 0$, and
\[
[x]=1
\]
means $x_s=0$ for $s>0$.

We use $x^{(s)}_i$ to denote the $i$-th Chern class of
$\mathcal{S}_s/\mathcal{S}_{s-1}$ for $s=1,\cdots, n+1$, where
$\mathcal{S}_0=0, \mathcal{S}_{n+1}=\mathcal{O}^{\oplus N}$. The
quantum cohomology of $F(k_1,k_2,\cdots,k_n,N)$ is generated by all
$x^{(s)}_{i_s}$ with $s=1,\cdots,n+1$ and $i_s =1,\cdots, k_s -
k_{s-1}$, i.e. the quantum cohomology
of $F(k_1,k_2,\cdots,k_n,N)$ has the form
\begin{equation}\label{qc1}
\mathbb{C}[x^{(1)}_1, \cdots, x^{(1)}_{k_1}, x^{(2)}_1, \cdots,
x^{(2)}_{k_2-k_1},\cdots, x^{(n+1)}_1, \cdots,
x^{(n+1)}_{N-k_{n}}]/I_q.
\end{equation}
The ideal of relations $I_q$ is generated by the coefficients of
$\lambda$ in the polynomial  \cite{DS}
\[
\lambda^N - \det(H+\lambda I),
\]
where $H$ is the $N\times N$ matrix \\ \\
\makebox[\textwidth][c]{ $\left(
\begin{array}{cccccccccccc}
x^{(1)}_1 & \cdots & x^{(1)}_{k_1} & 0 & \cdots & -(-1)^{k_2-k1} q_1
&
\cdots & 0 & \cdots & \cdots & 0 & 0 \\
-1 & \cdots & 0 & 0 & \cdots & 0 & \cdots & 0 & \cdots & \cdots & 0
& 0 \\
\vdots & \ddots & \vdots & \vdots & \ddots & \vdots & \ddots &
\vdots & \ddots & \ddots & \vdots & \vdots \\
0 & \cdots & -1 & x^{(2)}_1 & \cdots & x^{(2)}_{k_2-k_1} & \cdots &
-(-1)^{k_3-k-2}q_2 & \cdots & \cdots & 0 & 0 \\
0 & \cdots & 0 & -1 & \cdots & 0 & \ddots &
 & \ddots & \ddots & \vdots & \vdots \\
\vdots & \ddots & \vdots & \vdots & \ddots & \vdots & \ddots &
 & \ddots & \ddots & \vdots & \vdots \\
0 & \cdots & 0 & 0 & \cdots & 0 & \cdots & \cdots & x_1^{(n+1)} &
\cdots & x^{(n+1)}_{N-k_n-1} & x^{(n+1)}_{N-k_n} \\
0 & \cdots & 0 & 0 & \cdots & 0 & \cdots & \cdots & -1 & \cdots & 0
& 0 \\
\vdots & \ddots & \vdots & \vdots & \ddots & \vdots & \ddots &
\vdots & \ddots & \ddots & \vdots & \vdots \\
0 & \cdots & 0 & 0 & \cdots & 0 & \cdots & \cdots & 0 & \cdots & -1
& 0 \\
\end{array} \right)$
}\\ \\ \\
For any integer $l$ such that $1 \leq l \leq n$ and $n+1$ variables
$X_1,\cdots,X_{n+1}$, define
\[
\tau_{l}(X_1 X_2 \cdots X_{n+1}) = \left.(X_1 X_2 \cdots
X_{n+1})\right|_{X_l X_{l+1} = -q_l}.
\]
Expanding the determinant along the lines shows
\[
\det(H+\lambda I) = \sum_{l_a < l_{a+1}-1 \atop 0 \leq s \leq s(n)}
\tau_{l_1}\cdots \tau_{l_s}(Q_1 Q_2 \cdots Q_{n+1})
\]
up to signs that can be absorbed through redefinition of the
$q_l$'s, where
\[
Q_s = \sum_{i=0}^{k_s-k_{s-1}} \lambda^{k_s-k_{s-1}-i} x^{(s)}_{i},
\]
$s(n) = n/2$ for even $n$ and $s(n) = (n+1)/2$ for odd $n$.
By setting
$\lambda=1$, we see the relations in $I_q$ are given by
\begin{equation}\label{(2,2)relation}
\sum_{l_a < l_{a+1}-1 \atop 0 \leq s \leq s(n)} \tau_{l_1}\cdots
\tau_{l_s}([x^{(1)}][x^{(2)}] \cdots [x^{(n+1)}]) = 1
\end{equation}
with $\mathrm{deg}(q_s) = k_{s+1} - k_{s-1}$. In the classical
limit, i.e. $q_s \to 0$ for all $s=1,\cdots,n$,
\eqref{(2,2)relation} reduces to the relation of classical
cohomology of $F(k_1,k_2,\cdots,k_n,N)$:
\[
[x^{(1)}][x^{(2)}] \cdots [x^{(n+1)}] = 1.
\]

Now we give another representation of the quantum cohomology of
$F(k_1,k_2,\cdots,k_n,N)$, which has a form more suitable for (0,2)
generalization. First we extend the set of generators by allowing
the subscript of $x^{(s)}_i$ to take all positive integers. We claim
that \eqref{qc1} is isomorphic to
\begin{equation}\label{qc2}
\mathbb{C}[x^{(1)}_1, x^{(1)}_2 \cdots, x^{(2)}_1,\cdots,
x^{(n+1)}_1, \cdots]/(I+R),
\end{equation}
where $I$ is generated by the homogeneous components of
\begin{equation}\label{xxxx=1}
[x^{(1)}][x^{(2)}] \cdots [x^{(n+1)}] - 1
\end{equation}
and $R$ is generated by $\tilde{x}^{(s)}_{i_s}, s=1,\cdots,n+1, i_s
> k_s - k_{s-1}$. Here $\tilde{x}^{(1)}_r = x^{(1)}_r$ and
\begin{equation}\label{tilde:x}
\tilde{x}^{(s)}_r = \left\{
\begin{array}{ll}
x^{(s)}_r, & r<k_s-k_{s-2}, \\
x^{(s)}_r + q_{s-1} y^{(s-1)}_{r-k_s+k_{s-2}}, & r \geq k_s-k_{s-2}
\end{array} \right.
\end{equation}
for $s=2,\cdots,n+1$ and $y^{(s)}_r$ is defined to satisfy
\begin{equation}\label{xy}
[x^{(s)}][y^{(s)}]=1.
\end{equation}
\eqref{qc1} and \eqref{qc2} are obviously equivalent in the
classical limit. Now we show these two representations are equivalent
for arbitrary $q_s$.

By definition,
\[
[x^{(s)}] = [\tilde{x}^{(s)}] - q_{s-1} [y^{(s-1)}],~2 \leq s \leq
n+1.
\]
Plugging the expression above in the relation \eqref{xxxx=1}, we get
\[
[x^{(1)}]~([\tilde{x}^{(2)}] - q_1 [y^{(1)}])~([\tilde{x}^{(3)}] - q_2
[y^{(2)}]) \cdots ([\tilde{x}^{(n+1)}] - q_n [y^{(n)}]) = 1.
\]
By using the definition of $[y^{(s)}]$ \eqref{xy}, the left hand
side of the identity above can be shown to be equal to
\[
\sum_{l_a < l_{a+1}-1 \atop 0 \leq s \leq s(n)} \tau_{l_1}\cdots
\tau_{l_s}([\tilde{x}^{(1)}][\tilde{x}^{(2)}] \cdots
[\tilde{x}^{(n+1)}]).
\]
Comparing to \eqref{(2,2)relation}, we see \eqref{qc2} is isomorphic
to \eqref{qc1} for arbitrary $q_s$. We will see the quantum sheaf
cohomology has the same form as \eqref{qc2} with the ideal $R$
deformed. By the argument above we have shown that, on the (2,2) locus, the quantum
sheaf cohomology reduces to the ordinary quantum cohomology
\eqref{qc1}.

\subsection{Deformed tangent bundle}

The deformed tangent bundle of $F(k_1,k_2,\cdots,k_n,N)$ can be
described by turning on (0,2) deformations of the quiver GLSM. Again, these
deformations are encoded in the $E$-terms. Denote by
$\Lambda_{i,i+1}$ the Fermi multiplet corresponding to the chiral
multiplet $\Phi_{i,i+1}$, i.e. $\Lambda_{i,i+1}$ and $\Phi_{i,i+1}$
combine to give the (2,2) chiral multiplet when the deformations are
turned off. Up to field redefinitions, the $E$-terms with the
most general linear deformations are given by
\begin{equation}\label{E_F}
\begin{split}
&\overline{D}_+ \Lambda_{s,s+1}= \Phi_{s,s+1} \Sigma^{(s)} -
\Sigma^{(s+1)} \Phi_{s,s+1} + \sum_{t=1}^n u^s_t ({\rm Tr}\Sigma^{(t)})
\Phi_{s,s+1},\\
&\quad s=1,\cdots,n-1\\
&\overline{D}_+ \Lambda_{n,n+1}^i= \Phi_{n,n+1} \Sigma^{(n)} + ({\rm
Tr}\Sigma^{(t)}) {A_t}_j^i \Phi_{n,n+1}^j, ~i,j=1,\cdots,N,
\end{split}
\end{equation}
where we have suppressed the gauge indices. $\Sigma_s$ is the chiral
multiplet in the adjoint representation of $U(k_s)$. The parameters
$u^s_t$ are constants and $A_t$ are $N \times N$ matrices. When
$A_t=u^s_t=0$, we recover the $N=(2,2)$ theory.

The deformed tangent bundle $\mathcal{E}$ is defined by the
following short exact sequence
\begin{equation}\label{ses_Fd}
0 \to \bigoplus_{i=1}^n \mathcal{S}^\vee_i \otimes \mathcal{S}_i
\stackrel{g'}{\to} \bigoplus_{i=1}^n \mathcal{S}^\vee_i \otimes
\mathcal{S}_{i+1} \to \mathcal{E} \to 0.
\end{equation}
Comparing to \eqref{ses_F}, we see $g'-g$ is given by
\[
(\sigma^{(1)},\sigma^{(2)},\cdots,\sigma^{(n)}) \to (u^1_t {\rm
Tr}\sigma^{(t)} \Phi_{12}, u^2_t {\rm Tr}\sigma^{(t)}
\Phi_{23},\cdots,{\rm Tr}\sigma^{(t)} {A_t} \Phi_{n,n+1}).
\]

Generally speaking, the dimension of the moduli space of (0,2)
deformation is not the same as the number of parameters in the
$E$-terms. The actual dimension of the moduli space is given by
the dimension of the cohomology group $H^1(X, \mathrm{End} \, TX)$ for
target space $X$. For example, the Grassmannian $G(k,N)$ has \footnote{This can be
computed by the Borel-Weil-Bott Theorem.}
$\mathrm{dim} H^1(G(k,N), \mathrm{End}\, TG(k,N)) = N^2-1$, but there are $N^2$ parameters encoded in the matrix $A$ in \eqref{E_G}.
Actually, the deformed tangent bundle is isomorphic to the tangent
bundle when $A$ is proportional to the identity matrix \cite{GLS1, GLS2}. This accounts
for the difference between the dimension of the moduli space and the
number of constants parameterizing the (0,2) deformation. This
difference also exists in the case of general flag manifolds.
When $X=F(k_1,k_2,\cdots,k_n,N)$, since there are $n(n-1)+nN^2$ parameters in \eqref{E_F}, one should expect
\[
\mathrm{dim} H^1(X, \mathrm{End} \, TX) \leq n(n-1)+nN^2.
\]
We leave an explicit computation of $H^1(X,
\mathrm{End} \, TX)$ to future work.

\subsection{Coulomb branch}

As in the previous examples, once we get the classical relations, the quantum relations can be obtained by using the equations of motion derived from the one-loop effective $J$-functions on the Coulomb branch.

At a generic point of the Coulomb branch, the gauge group is broken
to $\prod_{s=1}^{n} U(1)^{k_s}$. The massless degrees of freedom are
the diagonal entries of $\sigma^{(s)}, s=1,\cdots,n$. We denote by
$\sigma^{(s)}_a$ the $a$-th diagonal element of $\sigma^{(s)}$ and
$\mathrm{Tr}\sigma^{(s)} = \sum_{a=1}^{k_s} \sigma^{(s)}_a$. The
mass matrices are
\begin{equation}\label{mM}
\begin{split}
m^{(s)}_{a b} &= \sigma^{(s)}_a-\sigma^{(s+1)}_b + \sum_{t=1}^n
u_t^s
\mathrm{Tr}(\sigma^{(t)}),\\
s&=1,\cdots, n-1,~~a=1,\cdots,k_{s},~~b=1,\cdots,k_{s+1}, \\
M_a &= \sigma^{(n)}_a I + \sum_{t=1}^n (\mathrm{Tr}\sigma^{(t)})
A_t.
\end{split}
\end{equation}

On the (2,2) locus, we interpret $\sigma^{(s)}_a$ as the Chern roots
of $\mathcal{S}_s^\vee$. Since $x^{(s)}_r$ is the $r$-th Chern class
of $\mathcal{S}_s/\mathcal{S}_{s-1}$, we can write $x^{(s)}_r$ as
polynomials in $\sigma$:
\begin{equation}\label{x}
\begin{split}
x^{(1)}_r &= (-1)^r S_{(1^r)}(\sigma^{(1)}), \\
x^{(s)}_r &= \sum_{i+j=r} S_{(i)}(\sigma^{(s-1)})(-1)^j
S_{(1^j)}(\sigma^{(s)}),~s=2,\cdots, n, \\
x^{(n+1)}_r &= S_{(r)}(\sigma^{(n)}).
\end{split}
\end{equation}
For fixed $s \leq n$, the degree $r$ component of
$[x^{(1)}][x^{(2)}] \cdots [x^{(s)}]$ is $(-1)^r
S_{(1^r)}(\sigma^{(s)})$, which is the degree $r$ elementary
symmetric polynomial in $\sigma^{(s)}$ up to sign. Consequently, all
the polynomials invariant under the permutation of
$(\sigma^{(s)}_1,\cdots,\sigma^{(s)}_{k_s})$ for all $s$ can be
generated by $x^{(i)}, i=1,\cdots,n+1$.

Note that \eqref{xy} implies
\begin{equation}\label{y}
\begin{split}
y^{(1)}_r &= S_{(r)}(\sigma^{(1)}), \\
y^{(s)}_r &= \sum_{i+j=r} (-1)^i S_{(1^i)}(\sigma^{(s-1)}) S_{(j)}(\sigma^{(s)}),~s=2,\cdots, n, \\
y^{(n+1)}_r &= (-1)^r S_{(1^r)}(\sigma^{(n)}).
\end{split}
\end{equation}

According to \eqref{EOM_general}, the equations of motion on the Coulomb branch read
\begin{equation}\label{EOM}
\begin{split}
&\prod_{a=1}^{k_2} m^{(1)}_{\alpha a} = -q_1,~ \alpha=1,\cdots,k_1, \\
&\prod_{b=1}^{k_{s+1}} m^{(s)}_{\alpha b} = -q_s \prod_{a=1}^{k_{s-1}} m^{(s-1)}_{a \alpha},~ \alpha=1,\cdots,k_s,~s=2,\cdots,n-1, \\
&\det (M_\alpha) = -q_n \prod_{a=1}^{k_{n-1}} m^{(n-1)}_{a \alpha}.
\end{split}
\end{equation}

\subsection{Quantum sheaf cohomology}

The quantum sheaf cohomology of $F(k_1,k_2,\cdots,k_n)$ take the
form of \eqref{qc2}. As before, $I$ is independent of the
deformation and is generated by the homogeneous components of
\[
[x^{(1)}][x^{(2)}] \cdots [x^{(n+1)}] - 1.
\]
One can compute that $b_0$ defined in \eqref{b0} is given by
\[
b_0^a = k_{s+1} - k_{s-1},~\sum_{i=1}^{s-1}k_{i} \leq a \leq
\sum_{i=1}^{s}k_{i},~ s=1,\cdots,n,~ k_0=0,~ k_{n+1}=N.
\]
Let's denote by $\rho^s_{ab}$ the weight associated with
$m^{(s)}_{ab}, s=1,\cdots,n-1,$ and by $\rho^n_a$ the weight
associated with $M_a$. Then
\[
b_0 = \sum_{s=1}^{n-1} \sum_{a_s,b_s} \rho^s_{a_s b_s} + N \sum_a
\rho^n_a,
\]
which shows that $\eta$ lies in the cone spanned by all the weights
associated with the matter multiplets. Thus, from the discussion of
section \ref{section:method}, the generators of $R$ multiplied by
$\Delta^2$ can be written as polynomials in $m^{(s)}_{ab}$ and
$\det{M_a}$. Following the same spirit, we first find these
polynomials on the (2,2) locus and then extend them to the (0,2)
region by modifying the expressions of $m^{(s)}_{ab}$ and
$\det{M_a}$ accordingly.

On the $(2,2)$ locus, $A_t=u^s_t=0$ in \eqref{E_F} and \eqref{mM}.
The relations $x^{(1)}_r = 0$ for $r>k_1$ are merely the consequence
of the first equation of \eqref{x}. Define
\[
\begin{split}
\Delta^{(s)} &= \prod_{a<b} (\sigma^{(s)}_{a}-\sigma^{(s)}_{b}), \\
\theta^{(s)}_c &= \prod_{a<b \atop a,b \neq c}
(\sigma^{(s)}_{a}-\sigma^{(s)}_{b}).
\end{split}
\]
From \eqref{x} and
\[
\begin{split}
\det (M^{(2,2)}_a) &= \sigma_a^{(n)N}, \\
\prod_{b=1}^{k_{s+1}} m^{(s)(2,2)}_{ab} =
\prod_{b=1}^{k_{s+1}}(\sigma^{(s)}_{a}-\sigma^{(s+1)}_{b}) &=
\sum_{i=0}^{k_{s+1}} {\sigma^{(s)}_a}^{k_{s+1}-i} (-1)^i
S_{(1^i)}(\sigma^{(s+1)}),
\end{split}
\]
we get
\begin{equation}\label{relation0_F}
\begin{split}
\Delta^{(s)} x^{(s+1)}_{k_{s+1}-k_{s}+r} &=
\sum_{a=1}^{k_{s}}(-1)^{a-1} {\sigma^{(s)}_a}^{r-1} \theta^{(s)}_a
\prod_{b=1}^{k_{s+1}}
m^{(s)(2,2)}_{ab},~~ s=1,\cdots,n-1, \\
\Delta^{(n)} x^{(n+1)}_{N-k_n+r} &= \sum_{a=1}^{k_n} (-1)^{a-1}(\det
(M^{(2,2)}_a)) {\sigma^{(n)}_a}^{r-1} \theta_a^{(n)}
\end{split}
\end{equation}
for $r > 0$, where we have used \eqref{xita}. From the formula
above, \eqref{EOM} and
\[
\prod_{a=1}^{k_{s}} m^{(s)(2,2)}_{ab} = (-1)^{k_s} \sum_{i=0}^{k_{s}}
{\sigma^{(s+1)}_b}^{k_{s}-i} (-1)^i S_{(1^i)}(\sigma^{(s)}),
\]
we obtain
\[
x^{(s+1)}_{k_{s+1}-k_s+r} = -q_s (-1)^{k_s-1} \sum_{i=0}^{k_{s-1}}
S_{(k_{s-1}-k_s+r-i)}(\sigma^{(s)})(-1)^i S_{(1^i)}(\sigma^{(s-1)}),
\]
for $s = 1, \cdots, n$, which recovers the (2,2) quantum relations
\eqref{tilde:x} from \eqref{y} up to signs that can be absorbed in $q_s$.

Now we turn on (0,2) deformation. We follow our previous
method and replace $m^{(s)(2,2)}_{ab}$ and $\det(M_a^{(2,2)})$ with
the expressions for $m^{(s)}_{ab}$ and $\det(M_a)$ given by \eqref{mM}.
Define $u^{(s)} = \sum_{t=1}^n u^s_t
\mathrm{Tr}(\sigma^{(t)})$, $s=1,\cdots,n-1$, and
\[
\begin{split}
\hat{\sigma}^{(s+1)}_a &= \sigma^{(s+1)}_a - u^{(s)}, \\
\check{\sigma}^{(s)}_a &= \sigma^{(s)}_a + u^{(s)}.
\end{split}
\]
From \eqref{mM}, we see $m^{(s)}_{ab} = \sigma^{(s)}_a -
\hat{\sigma}_b^{(s+1)} = \check{\sigma}^{(s)}_a - \sigma^{(s+1)}_b$.
Thus we have
\[
m^{(s)}_a \equiv  \prod_{b=1}^{k_{s+1}} m^{(s)}_{ab} =
\sum_{i=0}^{k_{s+1}} {\sigma_a^{(s)}}^{k_{s+1}-i} (-1)^i S_{(1^i)}
(\hat{\sigma}_b^{(s+1)}).
\]
Then in the generic case, \eqref{relation0_F} becomes
\[
\begin{split}
&\sum_{a=1}^{k_s}(-1)^{a-1} m_a^{(s)} {\sigma_a^{(s)}}^{r-1}
\theta_a^{(s)} \\
=& \Delta^{(s)} \sum_{i=0}^{\min\{k_{s+1},k_{s+1}-k_s+r\}}
(-1)^i S_{(1^i)}(\hat{\sigma}^{(s+1)}) S_{(k_{s+1}-k_s+r-i)}(\sigma^{(s)}).
\end{split}
\]
Therefore if we define
\[
R^{(s)}_r = \sum_{i+j=r} (-1)^i S_{(1^i)}(\hat{\sigma}^{(s)})
S_{(j)}(\sigma^{(s-1)}),~s=2,\cdots,n,
\]
then the (2,2) relations $x^{(s)}_{i_s}$ become $R^{(s)}_{i_s}$ in
the (0,2) case for $s=2,\cdots,n$ and $i_s > k_s - k_{s-1}$.

Another set of relations is obtained by modifying the second equation of \eqref{relation0_F}.
In the (0,2) case, since $\det (M_a) = \sum_{i=0}^N
\sigma^{(n)N-i}_a I_i$ with $I_i = I_i(\sum_{t=1}^n A_t
\mathrm{Tr}\sigma^{(t)})$, the right hand side of \eqref{relation0_F} becomes
\[
\sum_{i=0}^{\min\{N,N-k_n+r\}} I_i S_{(N-k_n+r-i)}(\sigma^{(n)})
\]
in the general case. If we define
\[
R^{(n+1)}_r = \sum_{i=0}^{\min\{N,r\}} I_i S_{(r-i)}(\sigma^{(n)})=
\sum_{i=0}^{\min\{N,r\}} I_i x^{(n+1)}_{r-i},
\]
then the (2,2) relations $x^{(n+1)}_{i}$ becomes $R^{(n+1)}_{i}$ in
the (0,2) case for $i > N - k_n$.

Quantum corrections can be incorporated by using \eqref{EOM} and
\[
\prod_{b=1}^{k_s} m^{(s)}_{ba} = (-1)^{k_s} \sum_{i=0}^{k_s}
{\sigma_a^{(s+1)}}^{k_s-i} (-1)^i S_{(1^i)}(\check{\sigma}^{(s)}).
\]
Now it is easy to see that if we define
\[
\begin{split}
&\tilde{R}^{(s)}_r  =  \left\{
\begin{array}{l}
R^{(s)}_r + q_{s-1} (-1)^{k_{s-2}} \sum\limits_{i=0}^{\min\{k_{s-2},
r-k_s+k_{s-2}\}}[ S_{(r-k_s+k_{s-2}-i)}(\sigma^{(s-1)})(-1)^i
S_{(1^i)}({\check{\sigma}}^{(s-2)})
], \\ \hfill r \geq k_s-k_{s-2}, \\
R^{(s)}_r, \hfill r < k_s-k_{s-2}
\end{array} \right.
\end{split}
\]
for $s=2,\cdots,n+1$,
then the quantum sheaf cohomology ring of $F(k_1,k_2,\cdots,k_n,N)$
with deformation given by \eqref{E_F} is
\begin{equation}\label{qsc}
\mathcal{A}_{F(k_1,k_2,\cdots,k_n,N)} = \mathbb{C}[x^{(1)}_1, x^{(1)}_2 \cdots, x^{(2)}_1,\cdots,
x^{(n+1)}_1, \cdots]/(I+R),
\end{equation}
where $I$ is generated by the homogeneous components of
\[
[x^{(1)}][x^{(2)}] \cdots [x^{(n+1)}] - 1
\]
and $R$ is generated by
\[
\{ x^{(1)}_{i_1}, \tilde{R}^{(s)}_{i_s} ~|~ i_1 > k_1, i_s > k_s -
k_{s-1}, s=2,\cdots,n+1 \}.
\]
Note that for any $t$ and $r$, $\tilde{R}^{(t)}_r$ is invariant
under the permutation of
$(\sigma^{(s)}_1,\cdots,\sigma^{(s)}_{k_s})$ for all $s$, thus from
the discussion below \eqref{x}, it is indeed a polynomial in the
generators.

\section{Dual deformation}\label{section:dual}

There is a biholomorphic duality between $F(k_1,k_2,\cdots,k_n,N)$
and $F(N-k_n,N-k_{n-1},\cdots,N-k_1,N)$ which maps $\mathcal{S}_i$
to $(\mathcal{O}^{\oplus N}/\mathcal{S}_{n+1-i})^\vee$. If we turn
on a (0,2) deformation on $F(k_1,k_2,\cdots,k_n,N)$ defined by
parameters $u^s_t$ and $A_t$, there should be a corresponding
deformation defined by $u'^s_t$ and $A'_t$, which determines the
same deformed tangent bundle. This leads to an IR duality between a
$U(k_1)\times U(k_2) \times \cdots U(k_n)$ GLSM and a
$U(N-k_n)\times U(N-k_{n-1}) \times \cdots U(N-k_1)$ GLSM with (0,2)
supersymmetry. The quiver diagrams of the two theories are
\[
\xymatrix@C=4pc{
*++[o][F]{k_1} & *++[o][F]{k_2} \ar@<-0.5ex>[l]_{\Phi_{12}} \ar@<0.5ex>@{-->}[l]^{\Lambda_{12}} & \cdots \ar@<-0.5ex>[l]_{\Phi_{23}} \ar@<0.5ex>@{-->}[l]^{\Lambda_{23}} & *++[o][F]{k_n} \ar@<-0.5ex>[l]_{\Phi_{n-1,n}} \ar@<0.5ex>@{-->}[l]^{\Lambda_{n-1,n}} & *++[F]{N} \ar@<-0.5ex>[l]_{\Phi_{n,n+1}} \ar@<0.5ex>@{-->}[l]^{\Lambda_{n,n+1}}
}
\]
and
\[
\xymatrix@C=4pc{
*++[o][F]{k'_1} & *++[o][F]{k'_2} \ar@<-0.5ex>[l]_{\Phi'_{12}} \ar@<0.5ex>@{-->}[l]^{\Lambda'_{12}} & \cdots \ar@<-0.5ex>[l]_{\Phi'_{23}} \ar@<0.5ex>@{-->}[l]^{\Lambda'_{23}} & *++[o][F]{k'_n} \ar@<-0.5ex>[l]_{\Phi'_{n-1,n}} \ar@<0.5ex>@{-->}[l]^{\Lambda'_{n-1,n}} & *++[F]{N} \ar@<-0.5ex>[l]_{\Phi'_{n,n+1}} \ar@<0.5ex>@{-->}[l]^{\Lambda'_{n,n+1}}
}
\]
where $k'_i = N - k_{n+1-i}$, solid arrows represent chiral multiplets and dashed arrows represent fermi multiplets.
In this section, we study the relationship between
$(u,A)$ and $(u',A')$ through the quantum sheaf cohomology. We start
with the duality of Grassmannians.

\subsection{Dual deformation of products of
Grassmannians}\label{section:dual_GG}

Suppose that the deformation on $G(k,N)$ is given by an $N \times
N$ matrix $A$ as in \eqref{E_G}. Under the duality, the short exact sequence \eqref{ses_G} dualizes to
\begin{equation}\label{ses_Q}
0 \to \mathcal{Q} \otimes \mathcal{Q}^\vee \to
\mathcal{Q} \otimes \mathcal{V}^\vee \to \mathcal{E}' \to 0
\end{equation}
on $G(N-k,N)$, which does not have a direct physical realization. But since $\mathcal{E}'$ is a deformed tangent bundle over $G(N-k,N)$, in general it should be equivalently described by the short exact sequence
\begin{equation}\label{ses_S}
0 \to \mathcal{S} \otimes \mathcal{S}^\vee {\to}
\mathcal{V} \otimes \mathcal{S}^\vee \to \mathcal{E}' \to 0
\end{equation}
with some deformation given by an $N \times N$ matrix $A'$. If $A$ and $A'$ define isomorphic deformed tangent bundles, the sheaf cohomology
rings must be isomorphic. The two rings are given by
\[
\mathcal{A}_{G(k,N)} =
\mathbb{C}[x_1,x_2,\cdots,y_1,y_2,\cdots]/(I+R),
\]
as in \eqref{ring0_G}, and
\[
\mathcal{A}_{G(N-k,N)} =
\mathbb{C}[x'_1,x'_2,\cdots,y'_1,y'_2,\cdots]/(I'+R')
\]
respectively. $I'$ is generated by the homogeneous components of
\[
[x'][y']-1,
\]
$R'$ is generated by $x'_i, i>N-k$, and
\[
R'_r = \sum_{i=0}^{\min\{N,r\}}I_i(y'_1 A')y'_{r-i},~ r>k.
\]

On the (2,2) locus, $x_i$ and $y_i$ are Chern classes of
$\mathcal{S}$ and $\mathcal{Q}$, the isomorphism between the rings
is thus given by the following
\[
\begin{split}
f: \mathcal{A}_{G(k,N)} &\to \mathcal{A}_{G(N-k,N)} \\
x_i &\mapsto (-1)^i y'_i \\
y_i &\mapsto (-1)^i x'_i.
\end{split}
\]
To make the notation succinct, we take the redefinition $x'_i
\to (-1)^i x'_i$ and $y'_i \to (-1)^i y'_i$. This does not change
the representation of the ring and the isomorphism now reads
\[
f(x_i)=y'_i, f(y_i)= x_i. \] Off the (2,2) locus, the isomorphism
becomes
\[
\begin{split}
f: \mathcal{A}_{G(k,N)} &\to \mathcal{A}_{G(N-k,N)} \\
x_i &\mapsto R'_i \\
R_i &\mapsto x'_i,
\end{split}
\]
i.e. $[x]=[R'], [R]=[x']$. We write
\[
I_i = I_i(y_1 A),~~ I'_i = I_i (y'_1 A').
\]
By definition
\[
[R]=[I][y],~~ [R']=[I'][y'],
\]
which yields
\begin{equation}\label{I->I'}
[I] = [I][x][y] = [R][R'] = [x'][I'][y'] = [I'].
\end{equation}
Consequently, the isomorphism of the rings suggests $y'_1 A' = y_1
A$ up to linear transformation of $\mathbb{C}^N$. Moreover,
\[
y_1 = -x_1 = -R'_1 = -I_1-y'_1 = - y_1 \mathrm{Tr} A -y'_1,
\]
which implies
\begin{equation}\label{A'(A)_G}
A' = - \frac{A}{1+\mathrm{Tr}A}.
\end{equation}
As a consistency check, we perform the duality twice and get
\[
A'' = - \frac{A'}{1+\rm{Tr}A'} = A.
\]
Note that quantum corrections do not change this result because
$[R] = [I][y] + q[y]$ implies
\[
[I]+q = [I']+q'
\]
and we can take $q=q'$, i.e. the two GLSMs in duality have the same FI
parameter and $\theta$-angle, and the $E$-terms are related
through \eqref{A'(A)_G}. The matrices with $\mathrm{Tr}A = -1$ seem to signal a new singular locus, however we expect the divergence of \eqref{A'(A)_G} along this locus is due to the failure of the assumption that the dual vector bundle can be described by the short exact sequence \eqref{ses_S} on $G(N-k,N)$. Along such locus, the dual bundle can only be described by \eqref{ses_Q}.

This method can be generalized to products of Grassmannians. Assume
that the deformation on $G(k_1,N_1)\times G(k_2,N_2)$ is given by
$N_1 \times N_1$ matrices $A, B$ and $N_2 \times N_2$ matrices $C,
D$ as in \eqref{E_GG}. The quantum sheaf cohomology is given by
\eqref{ring_GG}. On the other hand, the deformation on $G(N_1-k_1,N_1) \times
G(N_2-k_2,N_2)$ is given by $A',B',C',D'$. We take
\[
I_{1i} = I_i(y_{11} A + y_{21} B),~ I_{2i} = I_i(y_{11} C + y_{21}
D)
\]
and correspondingly on the dual side. The isomorphism is given by
\begin{equation}\label{iso_GG}
\begin{split}
&[x_i] = [R'_i] = [y'_i][I'_i]+q'_i[y'_i], \\
&[R_i] = [y_i][I_i]+q_i[y_i] = [x'_i],~~ i=1,2,
\end{split}
\end{equation}
from which we get
\[
[I_1]=[I'_1],~[I_2]=[I'_2]
\]
as \eqref{I->I'} by identifying $q_i$ with $q'_i$. This implies
\begin{equation}\label{A=A'_GG}
\begin{split}
y_{11} A + y_{21} B &= y'_{11} A' + y'_{21} B' ,\\
y_{11} C + y_{21} D &= y'_{11} C' + y'_{21} D'
\end{split}
\end{equation}
up to linear transformation of $\mathbb{C}^N$.
From the degree
one terms of \eqref{iso_GG} we get the equations
\[
\left\{
\begin{array}{c}
(1+\mathrm{Tr}(A)) x_{11} + \mathrm{Tr}(B) x_{21} = y'_{11}, \\
\mathrm{Tr}(C) x_{11} + (1+\mathrm{Tr}(D)) x_{21} = y'_{21},
\end{array}
\right.
\]
which can be solved to give
\begin{equation}\label{solution}
\left(
\begin{array}{c}
x_{11} \\ x_{21}
\end{array}
\right) = \frac{1}{m} \left(
\begin{array}{cc}
1+\mathrm{Tr}(D) & -\mathrm{Tr}(B) \\
-\mathrm{Tr}(C) & 1+\mathrm{Tr}(A)
\end{array} \right)
\left( \begin{array}{c} y'_{11} \\ y'_{21} \end{array} \right),
\end{equation}
where
\[
m = (1+\mathrm{Tr}(A))(1+\mathrm{Tr}(D))-\mathrm{Tr}(B)
\mathrm{Tr}(C).
\]
Because $y_{s1} = -x_{s1}$, \eqref{solution} means
\[
\begin{split}
y_{11} A + y_{21} B &= y'_{11} \left(- \frac{1+\mathrm{Tr}(D)}{m} A +
\frac{\mathrm{Tr}(C)}{m} B\right) + y'_{21}\left(\frac{\mathrm{Tr}(B)}{m} A -
\frac{1+\mathrm{Tr}(A)}{m} B\right), \\
y_{11} C + y_{21} D &= y'_{11}\left(- \frac{1+\mathrm{Tr}(D)}{m} C +
\frac{\mathrm{Tr}(C)}{m} D\right) + y'_{21}\left(\frac{\mathrm{Tr}(B)}{m} C -
\frac{1+\mathrm{Tr}(A)}{m} D\right).
\end{split}
\]
Then we see from \eqref{A=A'_GG} that the dual deformations are related by
\[
\left(
\begin{array}{cc} A' & B' \\ C' & D'
\end{array}
\right) = \frac{1}{m} \left(
\begin{array}{cc}
A & B \\
C & D
\end{array} \right)
\left( \begin{array}{cc} -1-\mathrm{Tr}(D) & \mathrm{Tr}(B) \\
\mathrm{Tr}(C) & -1-\mathrm{Tr}(A)
\end{array} \right).
\]

We can also take duality on only one of the Grassmannians. For
example, the same method leads to
\[
\left(
\begin{array}{cc} A' & B' \\ C' & D'
\end{array}
\right) = \left(
\begin{array}{cc}
-\frac{A}{1+\mathrm{Tr}(A)} & -\frac{\mathrm{Tr}(B)}{1+\mathrm{Tr}(A)}A + B \\
-\frac{C}{1+\mathrm{Tr}(A)} &
-\frac{\mathrm{Tr}(B)}{1+\mathrm{Tr}(A)}C + D
\end{array} \right)
\]
on $G(N_1-k_1,N_1)\times G(k_2,N_2)$, and
\[
\left(
\begin{array}{cc} A' & B' \\ C' & D'
\end{array}
\right) = \left(
\begin{array}{cc}
A-\frac{\mathrm{Tr}(C)}{1+\mathrm{Tr}(D)}B & -\frac{B}{1+\mathrm{Tr}(D)} \\
C-\frac{\mathrm{Tr}(C)}{1+\mathrm{Tr}(D)}D &
-\frac{D}{1+\mathrm{Tr}(D)}
\end{array} \right)
\]
on $G(k_1,N_1)\times G(N_2-k_2,N_2)$.

From \eqref{ring_Gs}, the same method can be generalized to
$G(k_1,N_1) \times G(k_2,N_2) \times \cdots \times G(k_n, N_n)$. We
just state the result for later use.

Assume the deformation on $G(k_1,N_1) \times G(k_2,N_2) \times
\cdots \times G(k_n, N_n)$ is given by the matrices $A_{st}$ as in
\eqref{E_GGGG}. Fix an integer $p$, $1 \leq p \leq n$, and take the
dual deformation on $G(N_1-k_1,N_1) \times \cdots \times
G(N_p-k_p,N_p) \times G(k_{p+1},N_{p+1}) \times \cdots \times G(k_n,
N_n)$ to be given by $A'_{st}$, then
\begin{equation}\label{A'(A)_GG}
\begin{split}
&A'_{si} = -\sum_{t=1}^p f_{ti} A_{st},~ 1 \leq i \leq p, \\
&A'_{sj} = A_{sj} - \sum_{t=1}^p g_{tj} A_{st},~ p+1 \leq j \leq n,
\end{split}
\end{equation}
where
\[
f=T^{-1}, g=T^{-1} W
\]
with
\[
\begin{split}
T_{ij}=\delta_{ij} + \mathrm{Tr}(A_{ij}),~ i,j=1,\cdots,p, \\
W_{ij}=\mathrm{Tr}(A_{ij}),~ i=1,\cdots,p,~ j=p+1,\cdots,m.
\end{split}
\]
Thus, starting from $G(k_1,N_1) \times G(k_2,N_2) \times
\cdots \times G(k_n, N_n)$, by taking dualities we get
\[
\sum_{i=0}^n {n \choose i} = 2^n
\]
different UV gauge theories dual to each other in the IR.

\subsection{Dual deformation of flag manifolds}

Now we study the duality on general flag manifolds. As mentioned at the beginning of this section, we want to find
the relationship between $(u^s_t,A_t)$ and
$({u'}^s_t,{A'}_t)$. We have seen the simplest situation, the
Grassmannians, in the last subsection. The key point was to derive
$[I]=[I']$ using the isomorphism between the quantum sheaf
cohomology rings.

When $u=u'=0$, we see from \eqref{qsc} that the ideal $R$ is
generated by $x^{(1)}_{i_1},
\tilde{x}^{(2)}_{i_2},\cdots,\tilde{x}^{(n)}_{i_n}$ and
$\tilde{R}^{(n+1)}_{i_{n+1}}$ for $i_s > k_s - k_{s-1}$. The
biholomorphic isomorphism identifies $\mathcal{S}_i$ on
$F(k_1,\cdots,k_n,N)$ with $(\mathcal{O}^{\oplus
N}/\mathcal{S}_{n+1-i})^\vee$ on $F(N-k_n,\cdots,N-k_1,N)$. Since
$[x^{(l)}]$ is interpreted as the total Chern class of
$\mathcal{S}_l/\mathcal{S}_{l-1}$, the isomorphism between the
quantum sheaf cohomology rings gives rise to the following
identities
\begin{equation}\label{x=x'}
\begin{split}
[x^{(1)}] &= [I'][{x'}^{(n+1)}] + q'_n (-1)^{N-k_2}[{y'}^{(n)}], \\
[x^{(2)}] + q_1 [y^{(1)}] &= [{x'}^{(n)}] + q'_{n-1}(-1)^{N-k_3}
[{y'}^{(n-1)}], \\
&\vdots \\
[x^{(n)}] + q_{n-1}(-1)^{k_{n-2}}[y^{(n-1)}] &= [{x'}^{(2)}] + q'_1
[{y'}^{(1)}], \\
[I][x^{(n+1)}] + q_n (-1)^{k_{n-1}} [y^{(n)}] &= [{x'}^{(1)}],
\end{split}
\end{equation}
where the prime indicates the corresponding quantities on the dual
side and we have used the fact that $[R^{(n+1)}]=[I][x^{(n+1)}]$. In
the classical limit, one can deduce
\begin{equation}\label{I=I'}
[I] = [I']
\end{equation}
from $[x^{(1)}][x^{(2)}]\cdots [x^{(n+1)}] = 1$. Actually, by taking
\[
q_1 = q'_n,~q_2(-1)^{k_1} =
q'_{n-1}(-1)^{N-k_3},~\cdots,q_n(-1)^{k_{n-1}} = q'_1,
\]
one can show that \eqref{I=I'} still holds even when quantum
corrections are taken into account. This implies
\begin{equation}\label{A=A'}
\sum_{j=1}^n A_j \mathrm{Tr}\sigma_j = \sum_{j=1}^n A'_j
\mathrm{Tr}\sigma'_j
\end{equation}
up to linear transformation of $\mathbb{C}^N$. Taking the degree
one terms on both sides of \eqref{x=x'}, we get
\begin{equation}\label{x_degree1}
\begin{split}
x^{(1)}_1 &= I_1 + {x'}^{(n+1)}_1, \\
x^{(s)}_1 &= {x'}^{(n+2-s)}_1,~s=2,\cdots,n, \\
I_1 + x^{(n+1)}_1 &= {x'}^{(1)}_1.
\end{split}
\end{equation}
Equation \eqref{x} yields
\[
\begin{split}
x^{(1)}_1 &= -\mathrm{Tr}\sigma_1, \\
x^{(s)}_1 &= \mathrm{Tr}\sigma_{s-1} -
\mathrm{Tr}\sigma_s,~s=2,\cdots,n, \\
x^{(n+1)}_1 &= \mathrm{Tr}\sigma_n,
\end{split}
\]
and
\[
\begin{split}
{x'}^{(1)}_1 &= -\mathrm{Tr}\sigma'_1, \\
{x'}^{(s)}_1 &= \mathrm{Tr}\sigma'_{s-1} -
\mathrm{Tr}\sigma'_s,~s=2,\cdots,n, \\
{x'}^{(n+1)}_1 &= \mathrm{Tr}\sigma'_n.
\end{split}
\]
Together with \eqref{x_degree1}, these identities give rise to the
relationship between $\mathrm{Tr}\sigma_a$ and
$\mathrm{Tr}\sigma'_a$
\[
\begin{split}
&\left(\begin{array}{ccccc}
-1-\mathrm{Tr}A_1 & -\mathrm{Tr}A_2 &
\cdots & \cdots & -\mathrm{Tr}A_n \\
1 & -1 & 0 & \cdots & 0 \\
0 & 1 & -1 & \cdots & 0 \\
\vdots & & & \ddots & \vdots \\
0 & \cdots & \cdots & 1 & -1
\end{array}\right)
\left(\begin{array}{c} \mathrm{Tr}\sigma_1 \\
\mathrm{Tr}\sigma_2 \\ \vdots \\ \vdots \\
\mathrm{Tr}\sigma_n
\end{array} \right) \\ = &
\left(\begin{array}{ccccc} 0 & 0 & \cdots & \cdots & 1 \\
0 & \cdots & \cdots & 1 & -1 \\
0 & \cdots & 1 & -1 & 0 \\
 & \cdots & \cdots & \cdots & \\
1 & -1 & 0 & \cdots & 0
\end{array} \right)
\left(\begin{array}{c} \mathrm{Tr}\sigma'_1 \\
\mathrm{Tr}\sigma'_2 \\ \vdots \\ \vdots \\
\mathrm{Tr}\sigma'_n
\end{array} \right)
\end{split}
\]
which enables us to express $\mathrm{Tr}\sigma_a$ in terms
of $\mathrm{Tr}\sigma'_a$ and vise versa. Then \eqref{A=A'} tells us
that the two deformations are related by
\begin{equation}\label{duality(u=0)}
A'_i = \left(1+\sum_{j=1}^n \mathrm{Tr} A_j \right)^{-1} \left[-\left(1+\sum_{j \neq n-i+1} \mathrm{Tr} A_j \right) A_{n+1-i} + \left(\mathrm{Tr} A_{n+1-i} \right) \sum_{j \neq n-i+1} A_j\right].
\end{equation}
Obviously, this result recovers \eqref{A'(A)_G} when $n=1$.

Actually \eqref{duality(u=0)} can be obtained from another approach
which generalizes to the case with nonzero $u$ and $u'$. We now turn
to this approach.

Consider the embedding
\begin{equation}\label{embed}
F(k_1,k_2,\cdots,k_n,N) \hookrightarrow G(k_1,N) \times \cdots \times
G(k_n,N)
\end{equation}
which sends a flag $V_{k_1} \subset \cdots \subset V_{k_n}$ to
$(V_{k_1}, \cdots, V_{k_n})$. Generally speaking, the
tangent bundles of $F(k_1,k_2,\cdots,k_n,N)$ and $G(k_1,N) \times
\cdots \times G(k_n,N)$ can be deformed independently. However, the two tangent bundles
can also be deformed in a coordinated fashion. Specifically, for any
deformed tangent bundle $\mathcal{E}$ of the flag manifold, there is
a deformed tangent bundle $\tilde{\mathcal{E}}$ of the ambient space such that the quotient $\tilde{\mathcal{E}}/\mathcal{E}$ is
isomorphic to the normal bundle $\mathcal{N}$ of the flag manifold
inside the product of Grassmannians. If $\tilde{\mathcal{E}}'$ is
the deformed tangent bundle over $G(N-k_n,N) \times \cdots \times
G(N-k_1,N)$ dual to $\tilde{\mathcal{E}}$, then the corresponding
deformed tangent bundle over $F(N-k_n,N-k_{n-1},\cdots,N-k_1,N)$
should be given by $\mathcal{E}'$ such that
$\tilde{\mathcal{E}}'/\mathcal{E}'$ is isomorphic to the normal
bundle $\mathcal{N}'$ of $F(N-k_n,N-k_{n-1},\cdots,N-k_1,N)$ inside
$G(N-k_n,N) \times \cdots \times G(N-k_1,N)$, i.e. we have the
following commutative diagram:
\begin{displaymath}
\xymatrix{
0\ar[r]&\mathcal{E}\ar[d]^{\cong}\ar[r]& \tilde{\mathcal{E}}\ar[d]^{\cong}\ar[r]&\mathcal{N} \ar[d]^{\cong}\ar[r]&0\\
0\ar[r]&\mathcal{E}'\ar[r]& \tilde{\mathcal{E}}'\ar[r]&
\mathcal{N}'\ar[r]&0 }
\end{displaymath}

The embedding \eqref{embed} can be realized physically by identifying
\[
{P_s}^i_{a_s} = {\Phi_{s,s+1}}_{a_s}^{a_{s+1}}
{\Phi_{s+1,s+2}}_{a_{s+1}}^{a_{s+2}} \cdots {\Phi_{n,n+1}}_{a_n}^i
\]
with the homogeneous coordinates on $G(k_s,N)$ for $s=1,\cdots,n$.
Each $P_s$ has a companion fermi multiplet $\Psi_s$, i.e. $P_s$ and
$\Psi_s$ combine to form a (2,2) chiral multiplet on the (2,2)
locus. It is easy to see
\[
\Psi_s = \sum_{l=s}^{n} \Phi_{n,n+1} \cdots \Phi_{l+1,l+2}
\Lambda_{l,l+1} \Phi_{l-1,l} \cdots \Phi_{s,s+1},
\]
from which we compute
\begin{equation}\label{E_embed}
\overline{D}_+ \Psi_s = P_s \Sigma_s + \left(\mathrm{Tr}\Sigma_t\right) \left( A^t + \sum_{a=s}^{n-1} u_a^t I \right) \Phi_s,
\end{equation}
where we have used \eqref{E_F}.

According to \eqref{E_GGGG}, \eqref{E_embed} tells us that the deformed tangent
bundle $\tilde{\mathcal{E}}$ of $G(k_1,N) \times \cdots \times
G(k_n,N)$ is given by the $N \times N$ matrices $A_{st}$ as in
\eqref{E_GGGG} with
\[
\begin{split}
A_{st} &= A_t + \sum_{a=s}^{n-1} u^a_t I, ~s \leq n-1, \\
A_{nt} &= A_t.
\end{split}
\]
Applying the above argument to $F(N-k_n,N-k_{n-1},\cdots,N-k_1,N)$,
we see the deformation of $G(N-k_n,N) \times \cdots \times
G(N-k_1,N)$ is given by the matrices $B_{st}$ with
\[
\begin{split}
B_{st} &= A'_t + \sum_{a=s}^{n-1} u'^a_t I, ~s \leq n-1, \\
B_{nt} &= A'_t.
\end{split}
\]
By switching the order, it is straightforward to see that the deformation of $G(N-k_1,N) \times
\cdots \times G(N-k_n,N)$ is given by the matrices $A'_{st}$ with
\[
\begin{split}
A'_{1t} &= A'_{n+1-t}, \\
A'_{st} &= A'_{n+1-t} + \sum_{t=n+1-s}^{n-1} {u'}^t_{n+1-t} I, ~s
\geq 2.
\end{split}
\]
Then the result of section \ref{section:dual_GG} shows
\begin{equation}\label{A'(A)_F'}
A'_{st} = - \sum_{l=1}^n A_{sl} M_{lt},
\end{equation}
where
\[
M = T^{-1}
\]
with $T$ defined by
\begin{equation}\label{T}
T_{ij} = \delta_{ij} + \mathrm{Tr}\left(A_j + \sum_{t=i}^{n-1} u^t_j I \right).
\end{equation}
Equivalently, if we define
\[
\begin{split}
X_{1t} &= -A_t-\sum_{l=1}^{n-1} u^l_t I, \\
X_{st} &= u^{s-1}_t, ~s = 2,\cdots,n,
\end{split}
\]
and
\[
\begin{split}
X'_{1t} &= A'_{n+1-t}, \\
X'_{st} &= u'^{n-s+1}_{n-t+1}, ~s = 2,\cdots,n,
\end{split}
\]
then \eqref{A'(A)_F'} can be written as
\begin{equation}\label{A'(A)_F}
X' = X T^{-1}.
\end{equation}
Equation \eqref{A'(A)_F} establishes the relationship between the deformation on
$F(N-k_n,N-k_{n-1},\cdots,N-k_1,N)$ and the deformation on
$F(k_1,k_2,\cdots,k_n,N)$. In physical language, this equation gives
the IR (A/2-twisted) duality between the corresponding (0,2) quiver
gauge theories. When $u=u'=0$, \eqref{A'(A)_F} reproduces
\eqref{duality(u=0)}.

For example, when $n=2$, \eqref{A'(A)_F} shows that the duality
between $F(k_1,k_2,N)$ and $F(N-k_2,N-k_1,N)$ is given by
\begin{equation}\label{n=2}
\begin{split}
&A'_1 = [(\mathrm{Tr}A_2+u^1_2 N)A_1-(1+\mathrm{Tr}A_1+u^1_1 N)A_2 -
(1+\mathrm{Tr}A_1) u^1_2 I + \mathrm{Tr}A_2 u^1_1 I]/m, \\
&A'_2 = [-(1+\mathrm{Tr}A_2)A_1 + (\mathrm{Tr}A_1)A_2 -
(1+\mathrm{Tr}A_2)u^1_1 I + \mathrm{Tr}A_1 u^1_2]/m, \\
& {u'}^1_1 = [(1 + \mathrm{Tr}A_1) u^1_2 - (\mathrm{Tr}A_2) u^1_1]/m, \\
& {u'}^1_2 = [(1 + \mathrm{Tr}A_2) u^1_1 - (\mathrm{Tr}A_1)
u^1_2]/m,
\end{split}
\end{equation}
where
\[
m = 1+u^1_1 N + (1 - u^1_2 N) \mathrm{Tr}A_1 + (1 + u^1_1 N)
\mathrm{Tr}A_2.
\]
As in the case of Grassmannians, the locus with non-invertible $T$ does not contribute a new component to the degenerate loci. However, along this locus the duality formula \eqref{A'(A)_F} is not applicable because the dual deformed tangent bundle cannot be described by a short exact sequence of the form \eqref{ses_Fd}.

\section{Conclusions}

In this paper we have proposed a method to compute the ring structure of quantum sheaf cohomology associated with (0,2) GLSMs with (2,2) locus. We applied this method to products of Grassmannians and flag manifolds, which are described by nonabelian gauge theories, and represented the quantum sheaf cohomology rings in terms of generators and relations \eqref{ring_Gs} \eqref{qsc}. We used our result to derive the dual deformations associated with the biholomorphic duality of flag manifolds \eqref{A'(A)_GG} \eqref{A'(A)_F}. Our description breaks down on codimension-one subvarieties of the moduli space of (0,2) deformations. These degenerate loci do not intersect the (2,2) locus.

There are a couple of open questions remaining. We did not compute the dimension of the moduli space of (0,2) deformation. The number of parameters we used to describe the deformation should be greater than the actual dimension of the moduli space in general. The ring structure of the quantum sheaf cohomology has not been proved mathematically even for Grassmannians. Such proof should in principle involve techniques of sheaf theory on Quot schemes. However, a purely mathematical derivation of the classical sheaf cohomology ring of flag manifolds should be a generalization of the derivation for Grassmannians in \cite{GLS2}. A mathematical proof of the dual deformation and the computation of quantum sheaf cohomology associated with (0,2) theories without (2,2) locus are left for the future.

\section*{Acknowledgements}

The author would like to thank Z. Lu, L. Mihalcea, S. Nawata and E. Sharpe for useful discussions and suggestions.

\end{document}